\documentstyle[aps,pre]{revtex}
\begin{document}
\title{Amplitude coda of classical waves in disordered media}
\author{Mauro S. Ferreira and Gerrit E. W. Bauer}
\address{Department of Applied Physics and \\
Delft Institute of Microelectronics and Submicron Technology, \\
Delft University of Technology, \\
Lorentzweg 1, 2628 CJ Delft, The Netherlands}
\date{\today} 
\maketitle

\begin{abstract}
The propagation of classical waves in the presence of a disordered medium 
is studied. We consider wave pulses containing a broad range of 
frequencies in terms of the configurationally averaged Green function of the system.
Damped oscillations in the time-dependent response trailing behind the 
direct arrival of the pulse (coda) are predicted, the periods of which 
are governed by the density of scatterers. 
\end{abstract}
\pagebreak

Wave propagation in disordered media is a broad research topic, with many applications. 
Electron transport in mesoscopic 
systems, light diffusion in opaque media, or acoustic propagation in the subsurface 
of the earth are examples of the interdisciplinary character 
of this general subject. All these examples share a 
number of common properties, the most important being that they are governed by 
wave equations. Methods and techniques used 
in one field of study can in principle be applied to others, and such an approach 
has proved successful in the past. Anderson localization\cite{Lee85} for 
instance, was originally discovered in the quantum realm of electrons but later 
understood to be a general wave phenomenon\cite{Wiersma97,Tourin97,Shengbook}. 

Other contributions from traditionally quantum methods into the study of classical 
waves are the diagrammatic perturbation technique\cite{Rossum99} and the random 
matrix theory\cite{Beenakker97}. Whereas the latter explains some universal features 
in the response of a disordered environment which are independent of the detailed 
structure of the system, the microscopic nature of the former allows a more 
specific modelling of the medium. If one wishes to image the medium by observing the 
signal which is generated by a source and scattered by the system inhomogeneities, 
the desired information must be sought in the non-universal part of the measurements. 

In addition to studies of classical wave propagation based on nearly monochromatic 
sources, the time-dependent response of a short pulse containing a 
broad range of frequencies is also of considerable interest. This is typically the 
case in seismic studies of the earth.
Excited either by earthquakes or by artificially controlled explosions,
seismic waves travel long distances before being registered by an array of
detectors. The measurements consist of time-dependent 
functions describing how the subsurface responds to the excitations. 
The portion of the wave lagging behind the first arrivals (the so-called 
coda) is delayed by being repeatedly scattered 
by interfaces and inhomogeneities and therefore carries information about 
the medium\cite{Snieder}. 
Only recently it has been realized that the coda of seismic waves can be used for 
mapping the subsurface of the earth as well as locating earthquake 
centres\cite{Sato}.

Unlike their quantum counterparts, the amplitude of classical waves 
can be probed quite trivially. This is the case for seismic and acoustic 
waves in general. Ultrasonic wave amplitudes are also observable and 
have been used to investigate multiple scattering effects in disordered 
media\cite{Scales,derode}. Amplitude and phase of electromagnetic waves have also 
been measured in the microwave region and used to treat dynamical aspects of the 
propagation\cite{Sebbah97,Sebbah00}. Such a wide applicability motivates us to 
study the time evolution of the wave amplitude within a disordered medium. 

Bearing in mind the microscopic approach mentioned above, we have recently 
investigated the time response of a disordered medium composed of small spheres 
(Rayleigh scatterers)\cite{Bauer2001}. The scatterer is characterized by a 
set of resonances which appear as poles of the scattering matrix on the lower 
half of the complex frequency-plane. A simple expression for the time-dependent wave 
amplitude is obtained when all but the lowest order poles are discarded. We 
subsequently considered an ensemble of randomly placed identical scatterers and related 
some features of the response to the microscopic details of the structure.

In this paper we extend this idea by accounting for all poles of the scattering 
matrix. On the one hand we loose the simplicity of the single pole approximation, 
but on the other the single scatterer contribution is better described by considering
scattering events previously neglected. In spite of the somewhat more complicated 
formalism, we are still able to relate the delayed-time response to the microscopic 
parameters of the system in the limit of impenetrable scatterers. Oscillations in 
the time-dependent amplitude of the wave are identified, the periods of which 
depend on the concentration of scatterers.

We consider a general scalar wave equation in the frequency-domain given by the 
following eigenvalue equation
\begin{equation}
\left\{-\nabla^2 + V({\vec r},\omega)\right\} \Psi_m({\vec r},\omega) = 
E_m(\omega) \Psi({\vec r},\omega) \,\,,
\label{eigen}
\end{equation}
where $\Psi_m({\vec r},\omega)$ is the wave field for a frequency $\omega$ at position 
${\vec r}$. Both the energy $E_m$ and the potential energy $V$ are frequency-dependent and 
given by $V({\vec r},\omega) = {\omega^2 \over c_0^2} 
\left(1 - {c_0^2 \over c^2({\vec r})}\right)$ and $E_m(\omega) = {\omega^2 \over c_0^2}$. 
The wave velocity $c({\vec r})$ is position-dependent, $c_0$ being the velocity for the 
homogeneous background medium. Associated with Eq. (\ref{eigen}) there is a 
Green function $G$ given by
\begin{equation}
G({\vec r},{\vec r_0},\omega) = \sum_m { \Psi_m({\vec r},\omega) 
\Psi^*_{m}({\vec r_0},\omega)\over {\omega^2 \over c_0^2} - E_m(\omega) + i \eta \, 
sgn(\omega) }\,\,,
\end{equation}
where $\eta$ is a positive infinitesimal number. $G({\vec r},{\vec r_0},\omega)$ 
represents the 
response measured at position ${\vec r}$ due to a stationary perturbation of frequency 
$\omega$ produced at ${\vec r_0}$. For the sake of simplicity we assume 
${\vec r}_0$ at the origin throughout this paper. The Fourier transform to 
time-domain gives the physically relevant quantity $G({\vec r},t)$, namely the response 
to a pulse containing the full spectrum of frequencies. From this quantity a simple 
convolution gives the wave amplitude for a pulse of arbitrary shape. 

For a disordered system with many scatterers we focus on the configurationally 
averaged Green function 
$\langle G({\vec r},t) \rangle$. In wave vector ${\vec k}$ and frequency space, 
it is given by 
\begin{equation}
\langle G({\vec k},\omega) \rangle = {1 \over  \left({\omega \over c_0}\right)^2 - k^2 - 
\Sigma({\vec k},\omega) }\,\,,
\label{Gav}
\end{equation}
where the so-called self-energy $\Sigma({\vec k},\omega)$ is a complex quantity. 
The solution to the disordered problem is determined 
by the behaviour of the self-energy. For spherically symmetric scattering cross 
sections ({\it s}-wave) and within the independent scatterer approximation\cite{Doniach}, 
valid for a low concentration of small defects $n$, the ${\vec k}$-dependence of the 
self-energy disappears and $\Sigma$ becomes
\begin{equation} 
\Sigma(\omega) = {n \; c_0 \, (S_0(\omega) - 1) \over 2 i \omega}\,\,, 
\label{Sigma}
\end{equation}
where $S_0(\omega)$ is the {\it s}-wave scattering matrix for the individual scatterers. 
Note that equations (\ref{Gav}) and (\ref{Sigma}) bridge the gap 
between the microscopic scale, in terms of $S_0$ of a single impurity, 
and the macroscopic medium, described by the average Green function $\langle G \rangle$. 

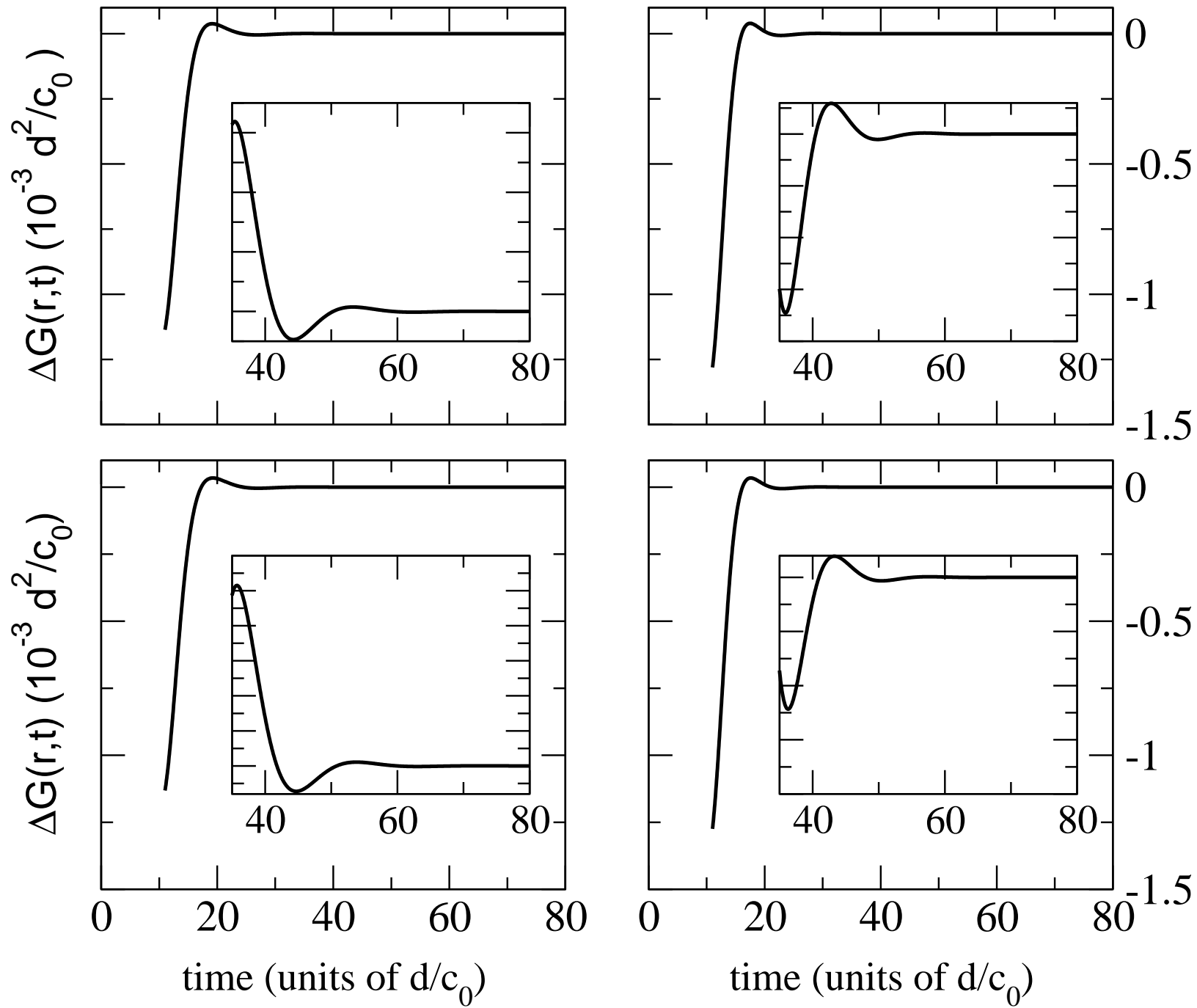
\begin{figure}
\caption{Scattered response $\langle \Delta G({\vec r},t) \rangle$ 
(in units of $10^{-3} \times d^2/c_0$) for $r = 10 \, d$. 
The graphs at the top correspond to the case of impenetrable spheres where 
$c=0$. The bottom graphs are for $c = c_0/5$. The graphs on the left and right 
are for $n = 5 \times 10^{-2} $(in units of $1/d^3$) and $n = 10^{-1} $(in units of 
$1/d^3$), respectively. The insets are amplified by a factor 10.}
\label{fig0}
\end{figure}

The scatterers are modelled by small spherical regions 
within which the waves propagate with velocity $c$. The corresponding {\it S}-matrix is 
given by\cite{Merzbacher}
\begin{equation}
S_0(\omega) = - e^{-i \omega d / c_0} \left ( 1 + {2 i \over 
{c_0 \over c} \cot\left({\omega d \over 2 c_0} \right ) - i } \right) \,\,,
\label{S0}
\end{equation}
where $d$ is the diameter of the scatterer. 
$S_0$ has an infinite number of poles 
in the frequency-domain, each one of them on the lower half of the complex 
frequency-plane. As shown in Ref. \onlinecite{Bauer2001}, a simple expression 
for the response is obtained when all but the lowest order poles are neglected. Here 
we want to include all poles of the {\it S}-matrix and investigate the effect this may bring 
to the final response. After substituting Eq. (\ref{S0}) into 
(\ref{Sigma}), the Fourier transform of Eq. (\ref{Gav}) 
yields the average Green function 
$\langle G({\vec r},t) \rangle$ in position and time domain. 
Alternatively, we can look at the scattered response 
$\langle \Delta G({\vec r},t) \rangle$, defined as the difference in amplitude 
between the total signal and the impurity-free response, which describes how the 
scattered portion of the waves evolves in time. Figure \ref{fig0} shows 
the calculated $\langle \Delta G({\vec r},t) \rangle$ for a random medium with identical 
scatterers at $n = 5 \times 10^{-2}$ and $n = 10^{-1}$ volume concentrations 
and $c = 0$ and $ c = {1 \over 5}$. $n$ and $c$ are expressed in 
units of $(1/d^3)$ and $c_0$, respectively, whereas the time is in units of $d/c_0$.
The distance $r$ to the source is arbitrarily fixed at $r = 10 d$.
As expected, the direct arrival of the wave is followed by an exponentially decaying 
response whose features depend on $n$ as well as on the scattering strength. 
In addition, clear oscillations in the scattered response can be seen in the 
insets of figure \ref{fig0}. Although not shown in the figure, oscillations are also 
present when $c > c_0$. 

To understand the origin of those oscillations it is helpful to simplify the problem
and assume that the wave velocity vanishes inside the scatterers, {\it i.e.}, $c = 0$. 
This is equivalent to a set of obstacles which are impenetrable to the waves, modelling 
for instance strong discontinuities in the constitutive parameters. 
One example is a solid porous medium filled with rarefied air, or any 
other environment with scatterers of relative low sound velocity. Alternatively, in the 
case of electromagnetic waves, obstacles with high refractive index are required.
With this simplification, the concentration $n$ and the diameter $d$ are the only 
variable parameters and the S-matrix becomes 
$S_0 = - e^{-i \omega d / c_0}$. The response $\langle G({\vec k},t) \rangle$ in 
wave-number and time domain is an intermediate step towards
$\langle G({\vec r},t) \rangle$ and results from the time Fourier transform of 
Eq. (\ref{Gav}), given by 
\begin{equation}
\langle G({\vec k},t) \rangle = {1 \over 2 \pi} \int_{-\infty}^{+\infty} d\omega 
\left( {\omega \over c_0} \right) 
\left\{ { e^{-i \omega t} \over \left( {\omega \over c_0} \right)^3 - 
			      \left( {\omega \over c_0} \right) k^2 + 
			      {n \over 2 i} \left[  e^{-i \omega d / c_0} + 1\right] } \right\}\,\,.
\label{t-Fourier}
\end{equation}
Although the spatial part of the Fourier transform remains to be done, 
Eq. (\ref{t-Fourier}) is useful for understanding the time dependence of the response. 
The frequency integral can be evaluated by contour integration. 
The poles are at $k$-dependent frequency 
values which govern the actual dispersion relation of the wave motion.
In the absence of scatterers ($n=0$) for instance, the poles are at 
$\omega = c_0 k$ and a further Fourier transform into space-domain gives 
the usual free-space solution\cite{morse} 
${\cal G}({\vec r},t) = {c_0 \over 4 \pi r} \delta(c_0 t - r)$. 
For $n \neq 0$, the poles acquire a concentration dependence $\omega_j(k,n)$, where $j$ 
is a labeling index. In general, the residue ${\cal R}$ associated with a first-order pole 
$\omega_j(k,n)$ is given by
\begin{equation}
{\cal R}[\omega_j(k,n)] = {\omega_j(k,n) \;\; e^{-i \omega_j(k,n) t} \over 
2 \pi c_0 \; \phi^\prime_j}\,\,,
\label{Residue}
\end{equation}
where $\phi^\prime_j$ stands for the $\omega$ derivative of $\phi = 
\left( {\omega \over c_0} \right)^3 - 
\left( {\omega \over c_0} \right) k^2 + 
{n \over 2 i} \left[  e^{-i \omega d / c_0} + 1\right]$ evaluated at $\omega = \omega_j(k,n)$.
The time dependence of the residue is entirely in the exponential 
$e^{-i \omega_j(k,n) t}$. The integral of Eq. (\ref{t-Fourier}), written as 
a sum of residues, is
\begin{equation}
\langle G({\vec k},t) \rangle = 
- {i \over c_0} \sum_j {\omega_j(k,n) \;
e^{-i \omega_j(k,n) t} \over \phi^\prime_j}\,\,.
\label{sum_residues}
\end{equation}
In this equation each term gives an exponentially damped 
oscillatory contribution, whose period and rate of decay depend on the real 
and imaginary parts of $\omega_j(k,n)$, respectively. Whether this time 
dependence persists in the final response depends on the spatial 
Fourier transform. 

After integrating over the angular variables, the remaining transform becomes
\begin{equation}
\langle G({\vec r},t) \rangle = \sum_j 
\int_{-\infty}^{+\infty} dk \; {\cal A}(k,r) \; e^{-i \omega_j(k,n) t}\,\,,
\label{r-Fourier}
\end{equation}
where ${\cal A}(k,r) = {i \, k^2 \sin(k r) \, 
\omega_j(k,n) \over c_0 \, r \, \phi^\prime_j}$. 
Except for the damping component induced by the imaginary part of the pole, 
the integrand in Eq. (\ref{r-Fourier}) is a periodic 
function of $t$, oscillating with a frequency given by the real part of 
$\omega_j(k,n)$ and with an amplitude governed by ${\cal A}(k,r)$. For large values of $t$ 
the exponential oscillates rapidly as a function of $k$ and the dominant contribution 
to the integral comes from the vicinity of $k$-points at which $\omega_j(k,n)$ is
stationary. In other words, the values of $k$ that effectively contribute 
to the integral are those satisfying the condition ${d \omega_j \over d k} = 0$.
Although each pole in Eq. (\ref{t-Fourier}) contributes 
with a different term in (\ref{r-Fourier}),
those meeting the stationary phase condition will dominate the signal. 

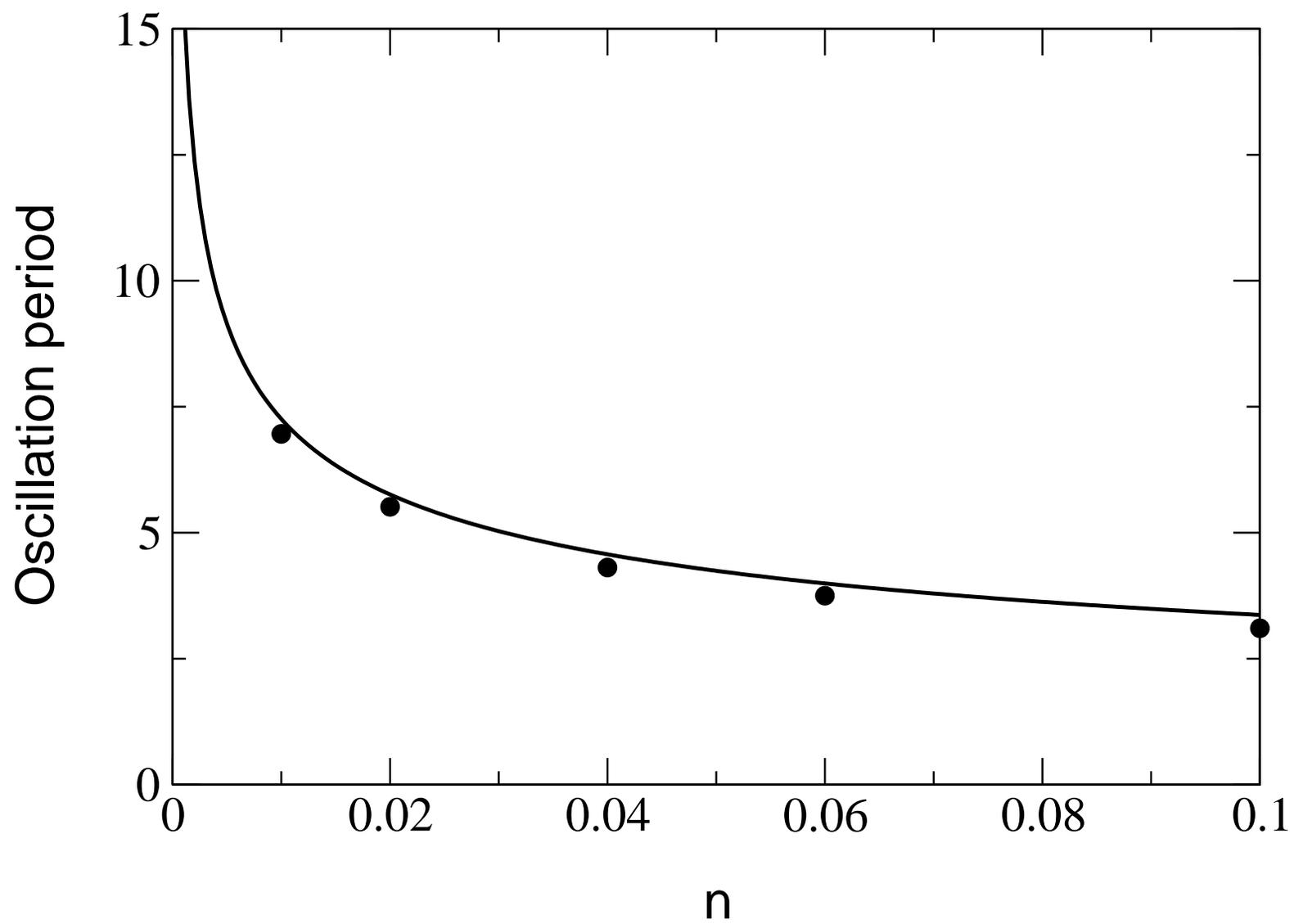
\begin{figure}
\caption{Oscillation periods (in units of $d/c_0$) as a function of the 
concentration $n$ (in units of $1/d^3$). 
The continuous line results from the stationary phase aproximation whereas the 
points are the periods obtained after numerical integration.}
\label{fig_periods}
\end{figure}

The existence of such stationary points is implied by the oscillations found 
in Fig. \ref{fig0} and we now try to locate them. A closer look at Eq. (\ref{t-Fourier}) 
shows that the poles $\omega_j(k,n)$ are functions of $k^2$. Therefore, $k = 0$ 
always satisfies the stationary phase condition ${d \omega_j \over d k} = 0$. We assume  
this value of $k$ as a possible candidate for explaining the observed results in 
Fig. \ref{fig0}. The corresponding poles are the solutions of a transcendental 
equation, which
can be further simplified if we expand the {\it S}-matrix for small $\omega$. 
Assuming that $n$ is small,
which is the basic condition for the validity of the independent scatterer approximation, 
the zero order term is sufficient and we find for the real part of the pole that 
$\Re[\omega_j(k=0,n)] = {c_0 \sqrt{3} \over 2}(n)^{1/3}$.
Associated with this pole is a simple expression for the oscillation period $T$ 
given by 
\begin{equation}
T = 2 \pi / \Re[\omega_j(k=0,n)] = \left({ 4 \pi \over c_0 \sqrt{3}}\right) 
\sqrt[3]{1 \over n}\,\,.
\label{T}
\end{equation}
The predicted oscillation periods based on the stationary 
phase argument, shown as a continuous line in figure \ref{fig_periods}, agree 
very well with the observed periods obtained by numerical integration 
of the Green function, represented by the points. Although the stationary phase 
approximation provides no information about the respective weight
of the oscillations, it is sufficient to explain their existence in 
the time response $\langle G({\vec r},t) \rangle$. With such a simple 
expression for the period, the oscillations are
easily related to a microscopic structural parameter of the
system, namely the average separation $a = 1/\sqrt[3]{n}$ between scatterers. 

A simple physical picture for the oscillations in the time-dependent amplitude 
is as follows. When scattered only once, the {\it S}-matrix $S_0$ induces a phase 
shift in the wave corresponding to a sign change. A second scattering event restores the 
original sign, which is again changed by a third scatterer, and so on. In this way, positive 
and negative contributions to the scattered response occur, depending on the 
number of scattering events. The maximum probability of being scattered a 
certain number of times depends on the time lapse, and therefore dictates the overall 
sign of the final response. For that reason the sign of the response changes as a 
function of time and we can also understand why the 
average distance between scatterers appears as the period of the oscillations.
On the opposite side of the velocity spectrum, infinitely large values of $c$ lead 
to $S_0 = e^{-i \omega d / c_0}$, causing no sign change in individual scattering events.
In this case, no oscillation should occur and, indeed, are not observed in the 
numerical solution for this limiting case. 

Therefore, according to Eq. (\ref{T}), it might be possible 
to obtain the values of $a$ and $n$, once such oscillations in 
$\langle G({\vec r},t) \rangle$ are observed. Whether this structural 
information can be extracted from an actual measurement depends on the type of 
disorder in the system. Within the conditions for the validity of the present 
model, {\it i.e.}, $\omega d \ll 1$ and $a/d \gg 1$, the oscillations do not 
depend on the distribution of the size of the scatterers.

Well defined oscillations in the time-delayed response have been recently observed in
ultrasonic waves transmitted across a 2-dimensional disordered structure\cite{derode}. 
They have a different origin, but we suspect that the superimposed beating pattern 
reflects the physics discussed here. This hypothesis could be tested by systematic 
experiments with different concentration of scatterers. 

In summary, we have observed oscillations in the time dependent response of a system 
composed of randomly placed spherical scatterers. Based on the stationary phase 
approximation, we have identified the origin of the oscillations and derived a 
simple expression for the periods in the limit of impenetrable scatterers, 
whose dependence on the mean separation between obstacles was established. 
It is suggested that such oscillations should be observable in disordered systems 
providing an indirect way of measuring the density of scatterers. 
When the scatterers are penetrable the simple analytical result presented here has to 
be modified, but can still be useful in the comparison to the numerical results.

We thank Cees Wapenaar for critically reading the manuscript. This work is part of 
the research program of the ``Stichting Technische Wetenschappen'' (STW) and the 
``Stichting Fundamenteel Onderzoek der Materie'' (FOM). G.B. acknowledges support 
by the NEDO program NAME.

\end{document}